\documentclass[12pt,a4paper]{article}
\usepackage{latexsym}
\usepackage{graphicx}
\usepackage{verbatim}
\usepackage{amssymb}

\catcode`\@=11
\def\marginnote#1{}
\newcount\hour
\newcount\minute
\newtoks\amorpm
\hour=\time\divide\hour by60
\minute=\time{\multiply\hour by60 \global\advance\minute by-\hour}
\edef\standardtime{{\ifnum\hour<12 \global\amorpm={am}%
        \else\global\amorpm={pm}\advance\hour by-12 \fi
        \ifnum\hour=0 \hour=12 \fi
        \number\hour:\ifnum\minute<10 0\fi\number\minute\the\amorpm}}
\edef\militarytime{\number\hour:\ifnum\minute<10 0\fi\number\minute}
\def\draftlabel#1{{\@bsphack\if@filesw {\let\thepage\relax
   \xdef\@gtempa{\write\@auxout{\string
      \newlabel{#1}{{\@currentlabel}{\thepage}}}}}\@gtempa
   \if@nobreak \ifvmode\nobreak\fi\fi\fi\@esphack}
        \gdef\@eqnlabel{#1}}
\def\@eqnlabel{}
\def\@vacuum{}
\def\draftmarginnote#1{\marginpar{\raggedright\scriptsize\tt#1}}
\def\draft{\oddsidemargin -.5truein
        \def\@oddfoot{\sl preliminary draft \hfil
        \rm\thepage\hfil\sl\today\quad\militarytime}
        \let\@evenfoot\@oddfoot \overfullrule 3pt
        \let\label=\draftlabel
        \let\marginnote=\draftmarginnote
   \def\@eqnnum{(\theequation)\rlap{\kern\marginparsep\tt\@eqnlabel}%
\global\let\@eqnlabel\@vacuum}  }

\def\preprint{\twocolumn\sloppy\flushbottom\parindent 1em
        \leftmargini 2em\leftmarginv .5em\leftmarginvi .5em
        \oddsidemargin -.5in    \evensidemargin -.5in
        \columnsep 15mm \footheight 0pt
        \textwidth 250mmin      \topmargin  -.4in
        \headheight 12pt \topskip .4in
        \textheight 175mm
        \footskip 0pt
        \def\@oddhead{\thepage\hfil\addtocounter{page}{1}\thepage}
        \let\@evenhead\@oddhead \def\@oddfoot{} \def\@evenfoot{} }

\def\titlepage{\@restonecolfalse\if@twocolumn\@restonecoltrue\onecolumn
     \else \newpage \fi \thispagestyle{empty}\c@page\z@ 
        \def\thefootnote{\fnsymbol{footnote}} }

\def\endtitlepage{\if@restonecol\twocolumn \else  \fi
        \def\thefootnote{\arabic{footnote}}
        \setcounter{footnote}{0}}  %\c@footnote\z@ }

\catcode`@=12
\relax
\def\bea{\begin{array}}
\def\bem{\begin{displaymath}}
\def\beq{\begin{equation}}

\def\eea{\end{array}}
\def\eem{\end{displaymath}}
\def\eeq{\end{equation}}
\def\Im{\mathop{\rm Im}}

\def\ov{\overline}

\def\Re{\mathop{\rm Re}}

\def\s2w{\sin^2 \theta_W}
\def\Tr{\mathop{\rm Tr}}

\relax
\def\dalpha{{\dot\alpha}}

\def\crbig{\\\noalign{\vspace {3mm}}}
\def\bigint{{\displaystyle\int}}

\def\Fint{\bigint d^2\theta\,}

\def\Dint{\bigint d^2\theta d^2\ov\theta\,}
\relax
\begin{document}
\topmargin-2.4cm
%\draft
%\preprint
%
%
%
%
\begin{titlepage}
\begin{flushright}
%hep--th/yymmnnn \\
\today
\end{flushright}
\vspace{1.3cm}

\begin{center}{\Large\bf
On Supercurrent Superfields and Fayet-Iliopoulos \\
\vspace{.4cm} Terms in \boldmath{$N=1$} Gauge Theories  }
\vspace{1.3cm}

{\large\bf Daniel Arnold$^1$, Jean-Pierre Derendinger$^1$ and Jelle Hartong$^2$}

\vspace{.8cm}
$^1$ Albert Einstein Center for Fundamental Physics, \\
Institute for Theoretical Physics, Bern University \\
Sidlerstrasse 5, CH--3012 Bern, Switzerland \\
\vspace{7mm}
$^2$ Niels Bohr Institute, Blegdamsvej 17, \\
DK--2100 Copenhagen \O, Denmark
\end{center}
\vspace{1.3cm}

\begin{center}
{\large\bf Abstract}
\begin{quote}
We revisit the supermultiplet structure of Noether currents for $N=1$
supersymmetric gauge theories. Using superfield identities and the field
equations we show how to derive a superfield equation for the divergences
of the Noether currents in terms of the supercurrent and anomaly
superfields containing 16$_B$+16$_F$ components. We refer to this as the
natural supercurrent structure as it is invariant under all local
symmetries of the theory. It corresponds to the ${\cal S}$--multiplet of
Komargodski and Seiberg. We clarify the on/off-shell nature of the
currents appearing in this multiplet and we study in detail the effect of
specific improvement transformations leading to 1) a Ferrara-Zumino
multiplet and to 2) a multiplet containing the new improved
energy-momentum tensor of Callan, Coleman and Jackiw. Our methods also apply
to supersymmetric gauge theories with a Fayet-Iliopoulos term. We
construct the natural supercurrent multiplet for such a theory and show how
to improve this to a formally gauge-invariant Ferrara-Zumino multiplet by
introducing a non-dynamical chiral superfield $S$ to ensure superfield gauge
invariance. Finally we study the coupling of
this theory to supergravity and show that $S$ remains non-dynamical if
the theory is $R$--symmetric and that $S$ becomes propagating if the
theory is not $R$--symmetric, leading to non-minimal 16$_B$+16$_F$
supergravity.
\end{quote}
\end{center}

\end{titlepage}
\renewcommand{\theequation}{\arabic{section}.\arabic{equation}}
\setcounter{footnote}{0}
\setcounter{page}{0}
\setlength{\baselineskip}{.6cm}
\setlength{\parskip}{.2cm}
\newpage
%
% BODY
%

\section{Introduction}

Any Poincar\'e-invariant $N=1$ supersymmetric theory has conserved supercurrent and 
energy-momentum tensor. In 1975, studying the Wess-Zumino model, Ferrara and Zumino found 
that these conserved Noether currents belong to a supermultiplet described by a 
real {\it supercurrent superfield} 
\beq
\label{SC0}
J_\mu = (\ov\sigma_\mu)^{\dalpha\alpha} \, J_{\alpha\dalpha},
\qquad\qquad
J_{\alpha\dalpha} = {1\over2}(\sigma^\mu)_{\alpha\dalpha} \, J_\mu,
\eeq
verifying a {\it supercurrent superfield equation} \cite{FZ}. Since a symmetric energy-momentum 
tensor and the 
supercurrent include $10_B+16_F$ fields\footnote{We use this notation for the number of bosonic and 
fermionic off-shell components.}, or $6_B+12_F$ fields with their conservation equations,
the supercurrent superfield necessarily includes other objects matching bosons and fermions.
Ferrara and Zumino also found that if
$
\ov D^\dalpha J_{\alpha\dalpha} = 0,
$
the theory has conserved dilatation and chiral $R$--symmetry currents and is actually superconformal
with $8_B+8_F$ operators: a symmetric, conserved and traceless energy-momentum tensor 
$\partial^\mu T_{\mu\nu}= {T^\mu}_\mu =0$, a conserved $R$--symmetry current, 
$\partial^\mu j_\mu=0$, 
and a conserved supercurrent with zero ``gamma trace", $\partial^\mu S_\mu = S_\mu\sigma^\mu =0$.
One can show on general grounds that these properties hold for every supersymmetric field 
theory, by demanding for instance that the conserved Noether charges derived from the currents 
generate the supersymmetry algebra \cite{SW, W}. The argument however assumes that the Noether 
charges are well-defined operators.

A generic supersymmetric theory is not superconformal and the chiral and dilatation 
currents are then not conserved. Their violations find sources in so-called classical or 
quantum anomalies, which can also, under the same circumstances, be described by 
anomaly superfields in the supercurrent superfield equation. This superfield equation should be such 
that the super-Poincar\'e conservation laws $\partial^\mu T_{\mu\nu} = \partial^\mu S_\mu =0$ 
are maintained, but $\partial^\mu j_\mu$, ${T^\mu}_\mu$, and $S_\mu\sigma^\mu$ may be nonzero.

In this work, we use the terminology {\it supercurrent structure} for the supercurrent superfield, the anomaly superfields and the corresponding supercurrent superfield equation. 

An exhaustive study of supercurrent structures in $N=1$ theories is a subtle problem which is not
our subject\footnote{See for instance refs.~\cite{MSW, Kuz, DS}.}. We consider in this article the already quite general situation where the supercurrent superfield equation is of the form
\beq
\label{SC1}
\ov D^\dalpha J_{\alpha\dalpha} = \Delta_\alpha, \qquad\qquad
D^\alpha J_{\alpha\dalpha} = - \ov \Delta_\dalpha ,
\eeq
with the complex spinor superfield $\Delta_\alpha$ which describes anomalies involving two 
contributions,\footnote{For consistency, the anomaly superfield $\Delta_\alpha$ is linear, 
$\ov{DD}\Delta_\alpha=0$. It then includes at most $24_B+24_F$ components. But the conservation
of the super-Poincar\'e currents reduces these numbers.} 
\beq
\label{SC2}
\Delta_\alpha = D_\alpha X + \chi_\alpha, \qquad\qquad
\ov\Delta_\dalpha = -\ov D_\dalpha\ov X + \ov\chi_\dalpha,
\eeq
where $X$ is a chiral superfield ($4_B+4_F$). The superfield $\chi_\alpha$ is also chiral, 
it includes an antisymmetric tensor verifying the Bianchi identity $D^\alpha \chi_\alpha
+ \ov D_\dalpha\ov\chi^\dalpha = 0$. Hence,
\beq
\label{SC3}
\chi_\alpha \,\,=\,\, -{1\over4} \, \ov{DD}D_\alpha \, U,
\qquad 
U^\dagger \,\,=\,\, U.
\eeq
and $\chi_\alpha$ also includes $4_B+4_F$ fields\footnote{Since $\chi_\alpha$ is invariant under $U\rightarrow U + \Lambda+\ov\Lambda$ with $\ov D_\dalpha \Lambda=0$}. In the supercurrent equation, 
$D_\alpha X$ and
$\chi_\alpha$ describe respectively the {\it chiral} and {\it linear} anomalies. 

For a given supersymmetric theory, one expects as a matter of principle to find 
an expression 
for the superfields $J_{\alpha\dalpha}$, $X$ and $\chi_\alpha$ in terms of the {\it off-shell} 
superfields of the theory. The use of superspace relies on linear supersymmetry, and then 
on off-shell fields. For on-shell fields solving the field equations, searching for superfield 
expressions for the currents does not make sense in general. 
The supercurrent equation, however, is a conservation equation which only holds 
for solutions of the field equations. 
This is a standard feature of the Noether prescription: currents can be directly calculated from the 
Lagrangian and the variational principle or the field equations provide then the (non-)conservation 
equations. 

The second remark at this stage is that the supercurrent structure of a given theory is not unique. 
There are superfield identities which act on $J_{\alpha\dalpha}$, $X$ and $\chi_\alpha$ leaving
the supercurrent equations (\ref{SC1}) and (\ref{SC2}) unchanged for the transformed superfields. Since identities do not contain significant, dynamical information, the transformation of the conserved $T_{\mu\nu}$ and $S_\mu$ is an
improvement of these currents. But other operators of the supercurrent structure are transformed 
significantly. For instance, the lowest component of $J_\mu$ is the current of a 
$U(1)_R$ rotation of the fields in the theory which is significantly modified in the transformation. 
Similarly, the relation between the divergence of the dilatation current and the trace of the 
energy-momentum tensor changes in the transformation. We will give a detailed discussion of these
transformations of a supercurrent structure.

In the next sections, we will proceed to give various 
expressions for $J_{\alpha\dalpha}$ for an arbitrary 
supersymmetric gauge theory with matter superfields. As indicated earlier, these expressions 
apply to off-shell superfields and are then unambiguous.
Then, secondly, using the field equations of the theory, we will calculate the anomaly superfields 
$X$ and $\chi_\alpha$ and equation (\ref{SC1}) will then contain all information on the 
conservation or violation of superconformal symmetries. This unambiguous procedure does 
work as easily for a theory with a Fayet-Iliopoulos term and, by construction, 
all expressions will be automatically gauge-invariant. 

The standard methods to derive a supercurrent structure are either\footnote{As described for 
instance in the textbooks \cite{GGRS} and \cite{BK}.}  to couple the theory to an off-shell (linearized)
background supergravity and obtain the currents from the variation of the background fields, or 
the (superconformal) superfield Noether procedure of ref.~\cite{MSW}. However, we should 
not rely on a specific 
off-shell supergravity formulation since we use improvement transformations which relate 
different supergravity formulations and we are primarily interested in theories which are neither 
scale- nor $R$-invariant. We then use a more heuristic but simpler method. 

Our $N=1$ superspace procedure is very similar to standard Noether currents and their 
(non-)conservation in a generic field theory. It is as
simple-minded as this: consider any function ${\cal L}(\varphi,\partial_\mu\varphi)$
of a single real field. A (linear) variation of $\varphi$ leads to the straightforward identity
\beq
\label{WZ6}
\partial^\mu\left( {\partial{\cal L}\over\partial\partial^\mu\varphi} \delta\varphi \right) =
\delta{\cal L} - \left( {\partial{\cal L}\over\partial\varphi} 
- \partial^\mu {\partial{\cal L}\over\partial\partial^\mu\varphi}\right) \delta\varphi.
\eeq
Then, if ${\cal L}$ is the Lagrangian and $\varphi$ solves the field equation, identity
(\ref{WZ6}) turns into a (non-)conservation equation for the Noether current
$$ 
j_\mu^N = {\partial{\cal L}\over\partial\partial^\mu\varphi} \delta\varphi 
$$
with source $\delta{\cal L}$. Notice however that the expression of the current can be derived
from ${\cal L}$ in terms of the off-shell field. 
Similarly, for all supersymmetric theories, one can define $J_{\alpha\dalpha}$, expressed in terms of
off-shell superfields, associated with a suitable identity which then generates the supercurrent equation, 
verified by on-shell fields only.

In Section \ref{sec2}, we briefly recall the field content of the supercurrent structure and equation, 
borrowing mostly from ref.~\cite{KS}, and we add some comments in preparation of the following sections.

Section \ref{sec3} considers a generic supersymmetric gauge theory, in three steps: deriving superfield
identities, obtaining a ``natural" supercurrent structure from these identities and improving the
natural structure. We discuss the nature of the energy-momentum tensor, its relation to
the dilatation current, the nature of the 
$R$--current and the role of auxiliary fields. We also identify the improvement transformation 
which leads to the supercurrent 
structure that contains the new improved energy-momentum tensor of Callan, Coleman and
Jackiw (CCJ) \cite{CCJ} and the Noether current of superconformal $U(1)_R$ transformations.
This particular supercurrent structure will be of importance in an
upcoming paper \cite{AADH}, in which we study the anomalies of $N=1$ super-Yang-Mills
in relation to the NSVZ $\beta$ function using effective field theory
techniques.

Section \ref{secFIterms} discusses the supercurrent structure of abelian gauge theories 
with a Fayet-Iliopoulos term, following the same steps as in section \ref{sec3}. 
The ``natural" supercurrent structure has $16_B+16_F$ operators, but we also show how to obtain a 
gauge-invariant Ferrara-Zumino $12_B+12_F$ structure with the formal introduction of $4_B+4_F$ 
new fields in a chiral superfield $S$ without any dynamical content. 
We then discuss the corresponding construction in supergravity. We point out that the chiral multiplet 
$S$ naturally appears through a gauge transformation in the superconformal formulation of new 
minimal Poincar\'e supergravity. We then study the old minimal version and we show how a generic
superpotential produces an obstruction which disappears in the global supersymmetry limit.
The obstruction is evaded if, as expected, the superpotential is $R$--symmetric and the
$R$--symmetry is gauged. With a generic superpotential, the way out is to turn $S$ into a dynamical
supermultiplet, {\it i.e.} to couple the globally supersymmetric theory to non-minimal $16_B+16_F$
supergravity. Our results provide us with a new perspective on the coupling of a
supersymmetric gauge theory with a Fayet-Iliopoulos term to supergravity
and complement the conclusions of ref.~\cite{KS}.
Conclusions and a technical appendix close the paper. 

\section{The supercurrent structure}\label{sec2}
\setcounter{equation}{0}

In preparation for the next sections, we begin with a discussion of the solution of the supercurrent 
superfield equation. We mostly follow earlier literature and in particular ref.~\cite{KS} but we 
also try to clarify and make precise several aspects which sometimes create confusion.
Again, the equations to solve are
\beq
\label{SC4}
\ov D^\dalpha\,J_{\alpha\dalpha} = D_\alpha X + \chi_\alpha, \qquad
\ov D_\dalpha X=0,\qquad
\chi_\alpha = -{1\over4}\ov{DD}D_\alpha\, U, \qquad U=U^\dagger.
\eeq
These equations for the supercurrent superfield are actually not the most general allowing conserved 
energy-momentum tensor and supercurrent \cite{MSW, Kuz, DS}, but they suffice for our purposes. 
In total, superfields $J_{\alpha\dalpha}$, $X$ and $\chi_\alpha$ include $40_B+40_F$ real components.
Since the supercurrent superfield equation is complex linear (it vanishes identically under $\ov{DD}$),
it imposes $2\times(12_B+12_F)$ conditions on the $40_B+40_F$ components to leave a 
solution expressed in terms of $16_B+16_F$ fields. 

The superfield identity
\beq
\label{impr1}
 2\,\ov D^\dalpha [ D_\alpha , \ov D_\dalpha ] \, {\cal G}
= D_\alpha \, \ov{DD}\,{\cal G} + 3\, \ov{DD} \, D_\alpha \,{\cal G},
\eeq
which holds for any superfield $\cal G$, can be used to transform the supercurrent structure 
into another solution of equations (\ref{SC4}):
\beq
\label{impr2}
\begin{array}{rcl}
J_{\alpha\dalpha} \qquad&\longrightarrow&\qquad \widetilde J_{\alpha\dalpha} = J_{\alpha\dalpha} 
+ 2\,[ D_\alpha , \ov D_\dalpha ] \, {\cal G} , 
\crbig
X \qquad&\longrightarrow&\qquad \widetilde X = X + \ov{DD}\,{\cal G} ,
\crbig
\chi_\alpha \qquad&\longrightarrow&\qquad \widetilde\chi_\alpha = 
\chi_\alpha + 3\, \ov{DD} \, D_\alpha \,{\cal G},
\end{array}
\eeq
with ${\cal G}$ real.
Hence, each theory admits in principle a (continuous) family of supercurrent structures. Notice that if 
${\cal G}$ is linear ($\ov{DD}{\cal G}= 0$), $\widetilde X=X$. Similarly, if ${\cal G}=\Psi + \ov \Psi$, 
$\ov D_\dalpha\Psi=0$, then $\widetilde\chi_\alpha = \chi_\alpha$. But the use of transformations 
(\ref{impr2}) may face various obstructions if conditions like gauge invariance or global definition are 
imposed on the supercurrent structure $J_{\alpha\dalpha}$, $X$, 
$\chi_\alpha$.\footnote{Although these superfields are not strictly speaking physical quantities.
These conditions 
have been discussed in ref.~\cite{KS} for specific theories. They will reappear in later sections.}

There are three obvious reductions. If $\chi_\alpha=0$, or if $\chi_\alpha$ can be canceled
using transformations (\ref{impr2}), the resulting supercurrent structure has $12_B+12_F$ fields and chiral 
anomaly $X$. 
This is the original Ferrara-Zumino \cite{FZ} structure. Transformations (\ref{impr2}) can still be used with
${\cal G}=\Psi+\ov\Psi$.
Similarly, if $X=0$, or if $X$ can be canceled using transformations (\ref{impr2}), the supercurrent structure
has again $12_B+12_F$ fields with linear anomaly $\chi_\alpha$. 
Transformations (\ref{impr2}) can still be used 
with a linear ${\cal G}$. Finally, if transformation (\ref{impr2}) can be used to obtain a supercurrent
with $X=\chi_\alpha=0$, it has $8_B+8_F$ fields and the theory is superconformal.

To solve in terms of component
fields the supercurrent equation, we use the following expansion of the chiral superfields 
$X$ and $\chi_\alpha$:
\beq
\label{4sc7}
\begin{array}{rcl}
X(y,\theta) &=& x + \sqrt2\,\theta\psi_X - \theta\theta\, f_X \, ,
\crbig
\chi_\alpha(y,\theta) &=& -i \lambda_\alpha + \theta_\alpha \, D 
+ {i\over2}(\theta\sigma^\mu\ov\sigma^\nu)_\alpha F_{\mu\nu}
- \theta\theta\, (\sigma^\mu\partial_\mu\ov\lambda)_\alpha,\end{array}
\eeq
in chiral coordinates\footnote{Appendix A gives complete formula.} and with 
$F_{\mu\nu}=\partial_\mu U_\nu - \partial_\nu U_\mu$. For the superfield $U$, the last eq.~(\ref{SC4}) implies
\beq
\label{4sc7b}
U = \theta\sigma^\mu\ov\theta\, U_\mu + i \,\theta\theta\ov{\theta\lambda} 
- i \, \ov{\theta\theta}\theta\lambda
+ {1\over2}\theta\theta\ov{\theta\theta}\, D + \ldots ,
\eeq
where the dots denote components of $U$ absent from $\chi_\alpha$.
The resulting supercurrent superfield can 
then be written\footnote{We do not define a normalization for the supercurrent $S_\mu$,
we will not use its explicit expression.}
\beq
\label{4sc8}
\begin{array}{rcl}
J_\mu (x,\theta,\ov\theta) &=& {8\over3}\, j_\mu(x) 
+ \theta (S_\mu + 2\sqrt2\, \sigma_\mu\ov\psi_X )
+ \ov\theta (\ov S_\mu - 2\sqrt2\, \ov\sigma_\mu\psi_X )
\crbig
&& - 2i \,\theta\theta\,\partial_\mu\ov x + 2i \,\ov{\theta\theta}\, \partial_\mu x
\crbig
&& \displaystyle
+ \theta\sigma^\nu\ov\theta\left[ 8\,T_{\mu\nu} - 4\,\eta_{\mu\nu}\Re f_X
- {1\over2}\epsilon_{\mu\nu\rho\sigma}\left( {8\over3}\,\partial^\rho j^\sigma - F^{\rho\sigma} \right)\right]
\crbig
&& \displaystyle - {i\over2} \theta\theta\ov\theta ( \partial_\nu S_\mu \sigma^\nu 
+ 2 \sqrt2\, \ov\sigma_\mu \sigma^\nu \partial_\nu\ov\psi_X)
\crbig
&& \displaystyle + {i\over2} \ov{\theta\theta}\theta ( \sigma^\nu \partial_\nu \ov S_\mu
+ 2\sqrt2\, \sigma_\mu \ov\sigma^\nu \partial_\nu\psi_X)
\crbig
&& \displaystyle
- {2\over3}\theta\theta\ov{\theta\theta}\, \Bigl(  2 \, \partial_\mu \partial^\nu j_\nu
- \Box j_\mu \Bigr)
\end{array}
\eeq
with $T_{\mu\nu} = T_{\nu\mu}$. This expression solves eq.~(\ref{SC4}) if $T_{\mu\nu}$
and $S_\mu$ verify conservation equations
\beq
\label{4sc9}
\partial^\mu T_{\mu\nu} = 0, \qquad\qquad
\partial^ \mu S_\mu = 0.
\eeq
Hence, $T_{\mu\nu}$ and $S_\mu$ will be (proportional to) the conserved energy-momentum 
tensor and the supercurrent. 

In addition, the supercurrent equation indicates that the following additional 
relations are verified:
\beq
\label{4sc10}
\begin{array}{c}
4 \, {T^\mu}_\mu = D + 6 \Re f_X, \qquad\qquad
\partial^\mu\,j_\mu = - {3\over2} \,\Im f_X, 
\crbig
(\sigma^\mu \ov S_\mu)_\alpha = 6\sqrt2\,\psi_{X\,\alpha} + 2i\,\lambda_\alpha.
\end{array}
\eeq
The first condition indicates that $D$ and $\Re f_X$ are sources for the 
trace of the energy-momentum tensor. Its precise significance depends on the specific 
energy-momentum tensor included in $J_\mu$: since $T_{\mu\nu}$ is defined up to improvements,
is it not in general true that a scale-invariant theory has a traceless energy-momentum tensor.
The second condition indicates that $\Im f_X$ induces the non-conservation of $j_\mu$, which
is related in general to an $R$ transformation acting in the theory. 
The third condition controls the violation of superconformal supersymmetry.

For $X\ne 0\ne\chi_\alpha$, the supercurrent superfield $J_\mu$ includes a conserved symmetric
energy-momentum tensor $T_{\mu\nu}$ 
($10_B - 4_B = 6_B$), the conserved supercurrent $S_\mu$ ($4\times (4-1)_F = 12_F$) 
and a vector current $j_\mu$ which is not conserved ($4_B$). Since conditions (\ref{4sc10})
eliminate $2_B+4_F$, the source superfields $X$ and $\chi_\alpha$ add $6_B+4_F$
fields, for a total of $16_B+16_F$ fields.

Denoting the components of $\cal G$ by $(C_g,\chi_g,v_{g\mu},\ldots)$, the component fields 
$j_\mu$, $S_\mu$ and $T_{\mu\nu}$ of $J_\mu$ change into
\beq
\label{WZ24}
\begin{array}{rcl}
j_\mu \qquad&\longrightarrow&\qquad \tilde j_\mu = j_\mu-3v_{g\mu}, 
\crbig
S_\mu \qquad&\longrightarrow&\qquad \tilde S_\mu = S_\mu + 8\sigma_{[\mu}\ov\sigma_{\nu]}\partial^\nu\chi_g,
\crbig
T_{\mu\nu} \qquad&\longrightarrow&\qquad \tilde T_{\mu\nu} = T_{\mu\nu} + (\partial_\mu\partial_\nu-\eta_{\mu\nu}\Box)C_g ,
\end{array}
\eeq
under the transformation (\ref{impr2}).
Clearly, $S_\mu$ and $T_{\mu\nu}$ are changed by improvements, {\it i.e.} 
trivially conserved terms whose $\mu=0$ components are spatial derivatives  
leaving the corresponding Noether charges unaffected. But, unless $v_{g\mu}=\partial^\nu A_{\mu\nu}$ 
with some $A_{\mu\nu}=-A_{\nu\mu}$, the vector field $j_\mu$ which is not in general conserved 
is more significantly transformed into a completely different current.
This new current could be associated, by Noether procedure,
to a different global transformation of the fields in the theory.

Some remarks are in order. 
Firstly, notice that  the components of the anomaly superfields $X$ and $\chi_\alpha$ 
appear in $J_\mu$.  Hence, the symmetric part of the $\theta\sigma^\nu\ov\theta$ component
of $J_\mu$ can only be identified with an energy-momentum tensor of the theory after subtraction of an anomaly contribution generated by $\Re f_X$, or by $D$, or by both,
since we may as well use the first eq.~(\ref{4sc10}) to modify the component expansion (\ref{4sc8}).\footnote{The omission of these anomaly contributions seems to be at the origin of the erroneous
no-go statement of ref.~\cite{CPS}, as observed in ref.~\cite{KS}.}

Secondly, even if, for a given theory, one expects to find expressions for $J_{\alpha\dalpha}$, 
$X$ and $\chi_\alpha$ in terms of superfields, {\it i.e.} in terms of off-shell fields,
equations (\ref{4sc8})--(\ref{4sc10}) only hold for on-shell fields. The interpretation of the components of
$J_\mu$ in terms of currents may require the field equations. This is in particular true, as we will see later 
on, for the auxiliary field contributions.

\section{Supersymmetric gauge theory} \label{sec3}
\setcounter{equation}{0}

In this section, we consider an arbitrary $N=1$ gauge theory with matter superfields $\phi^i$ in
some representation of the gauge group. In most expressions however, we will use the notation 
$\Phi$ to denote the collection of all chiral superfields $\phi^i$, viewed as a column matrix, and 
eliminate indices. Except otherwise indicated,
the gauge vector superfield $A$ is valued in this representation: $A=A^aT^a$, with generators
$T^a$ in the representation of $\Phi$.
Gauge transformations have a chiral parameter $\Lambda = \Lambda^aT^a$, 
$\ov D_\dalpha\Lambda=0$. They read:
\beq
\label{WZ1}
\begin{array}{rclrcl}
\Phi \quad&\longrightarrow&\quad e^\Lambda \, \Phi, \qquad&\qquad
\ov\Phi \quad&\longrightarrow&\quad \ov\Phi \, e^{\ov\Lambda},
\crbig
e^A \quad&\longrightarrow&\quad e^{-\ov\Lambda}\,e^A\,e^{-\Lambda}, \qquad&\qquad
e^{-A} \quad&\longrightarrow&\quad e^\Lambda \, e^{-A} \, e^{\ov\Lambda}
\end{array}
\eeq
and $\ov\Phi e^A \Phi$ is gauge-invariant. The gauge covariant supersymmetric derivatives are
\beq
\label{WZ2}
{\cal D}_\alpha\Phi = e^{-A} D_\alpha ( e^A \, \Phi ) ,
\qquad\qquad
\ov{\cal D}_\dalpha\ov\Phi = \ov D_\dalpha (\ov\Phi e^A) e^{-A}
\eeq
and $(\ov{\cal D}_\dalpha\ov\Phi) e^A ({\cal D}_\alpha\Phi)$ is then also gauge-invariant.\footnote{
As usual, $D_\alpha = \frac{\partial}{\partial\theta^\alpha}-i(\sigma^\mu\ov\theta)_\alpha \partial_\mu$
and
$\ov D_\dalpha = \frac{\partial}{\partial\ov\theta^\dalpha}-i (\theta\sigma^\mu)_\dalpha \partial_\mu$.}

\subsection{An identity for matter superfields}

We begin our discussion of the supercurrent with a superfield identity.
For any real function $K(\ov\Phi e^A\Phi)$, some simple manipulations lead to
\beq
\label{WZ3}
2\,\ov{D}^{\dot\alpha} \Bigl[(\ov{\cal D}_{\dalpha}\ov\Phi)K_{\Phi\ov\Phi}
({\cal D}_\alpha\Phi)\Bigr]
= -\ov{DD}D_\alpha K - 4 K_\Phi {\cal W}_\alpha\Phi - (\ov{DD} K_\Phi) ({\cal D}_\alpha\Phi),
\eeq
where
\beq
\label{WZ4}
{\cal W}_\alpha = -\frac{1}{4}\overline{DD}e^{-A}D_\alpha e^A,
\qquad\qquad
\ov {\cal W}_\dalpha = \frac{1}{4}DD e^A\ov D_\dalpha e^{-A} 
\eeq
are the non-abelian field strength superfields\footnote{Notice that with these standard but somewhat 
unfortunate definitions, $\ov{\cal W}_\dalpha = - ({\cal W}_\alpha)^\dagger$.} with gauge transformations
\beq
\label{WZ4b}
{\cal W}_\alpha \quad\longrightarrow\quad e^\Lambda {\cal W}_\alpha e^{-\Lambda}, 
\qquad\qquad
\ov {\cal W}_\dalpha \quad\longrightarrow\quad e^{-\ov\Lambda}\ov {\cal W}_\dalpha e^{\ov\Lambda}.
\eeq
The notation
\beq
\label{WZ5}
K_{\Phi} = \frac{\partial K}{\partial\Phi}, \qquad \qquad
K_{\ov\Phi} = \frac{\partial K}{\partial\ov\Phi}, \qquad\qquad
K_{\Phi\ov\Phi} = \frac{\partial^2 K}{\partial\Phi\partial\ov\Phi}
\eeq
is used.
We stress that the gauge-invariant 
eq.~(\ref{WZ3}) is an identity, it does not contain any information. 
It will be used to define the supercurrent superfield $J_{\alpha\dalpha}$ of a 
Wess-Zumino model with K\"ahler potential $K$, up to improvements to be discussed later on.

Notice that we use for simplicity the gauge-invariant variable $\ov\Phi e^A \Phi$. But formula 
(\ref{WZ3}) actually holds for an arbitrary gauge-invariant function $K$. We will also assume 
that the theory does not include gauge-singlet chiral superfields. Then, if the K\"ahler potential 
depends on real gauge-invariant variables like $\ov\Phi e^A\Phi$, it always 
has a global (non--$R$) $U(1)$ symmetry. 

\subsection{An identity for gauge superfields}

We may derive a similar identity for gauge superfields. The tool is the 
non-abelian Bianchi identity
\beq
\label{WZ10}
e^{-A}D^\alpha(e^A {\cal W}_\alpha e^{-A}) e^A = \ov D_\dalpha(e^{-A}\ov{\cal W}^\dalpha e^A).
\eeq
Multiplying (left) by ${\cal W}_\alpha$ and taking the trace gives
\beq
\label{WZ10b}
\ov D^\dalpha \Tr [ {\cal W}_\alpha\, e^{-A}\ov {\cal W}_\dalpha e^A  ]
= \Tr [ e^A {\cal W}_\alpha e^{-A} D^\beta ( e^A {\cal W}_\beta e^{-A} ) ].
\eeq
Then, for an arbitrary (gauge-invariant) holomorphic function $g(\Phi)$,
\beq
\label{WZ10c}
\begin{array}{rcl}
\ov D^\dalpha \Bigl[ (g+\ov g)\Tr [ {\cal W}_\alpha\, e^{-A}\ov {\cal W}_\dalpha e^A  ] \Bigr]
&=& (g+\ov g)\, \Tr [ e^A {\cal W}_\alpha e^{-A} D^\beta ( e^A {\cal W}_\beta e^{-A} ) ]
\crbig
&& + (\ov D^\dalpha\ov g)  \Tr [ {\cal W}_\alpha\, e^{-A}\ov {\cal W}_\dalpha e^A  ].
\end{array}
\eeq
Identities (\ref{WZ3}) and (\ref{WZ10c}) are the building blocks of the supercurrent structure for
the gauge-invariant Wess-Zumino model, which we consider next. 

\subsection{The theory and its supercurrent structures}

We consider the following gauge-invariant Wess-Zumino model:
\beq
\label{WZ7}
{\cal L}=\Dint K(\ov\Phi e^A\Phi)
+ \Fint \left[W(\Phi)+\frac{1}{4}g(\Phi)\widetilde{\Tr}({\cal W}^\alpha {\cal W}_\alpha)\right]+ {\rm h. c.}
\eeq
The holomorphic functions $W$ and $g$ are assumed invariant under the non-abelian 
supersymmetric gauge transformations (\ref{WZ1}). Gauge kinetic terms are normalized using
$\widetilde\Tr = {1\over T( R )}\Tr$, $\Tr(T^aT^b) = T( R )\delta^{ab}$. Notice that with our choice 
of variable, since\footnote{The prime denotes the first derivative of $K$ with respect to its 
variable $\ov\Phi e^A\Phi$.}
\beq
\label{WZ7b}
K_\Phi \Phi = \ov\Phi K_{\ov\Phi} = K^\prime \, \ov\Phi e^A\Phi,
\eeq
the K\"ahler potential part of the theory is always invariant under the non-$R$ $U(1)$ symmetry 
rotating all chiral superfields $\phi^i$ by the same phase. In general, a non-trivial gauge kinetic function
$g$ or a superpotential $W$ will break this symmetry. Assuming scale dimension $w$ for all 
chiral superfields\footnote{Our conventions for scale transformations are
$$
\begin{array}{rcl}
x^\mu \quad&\longrightarrow&\quad e^{-\lambda}\,x^\mu,
\\ 
\theta \quad&\longrightarrow&\quad e^{-\lambda/2}\,\theta,
\\
\Phi(x) \quad&\longrightarrow&\quad e^{w\lambda}\,\Phi(e^\lambda x).
\end{array}
$$}, scale invariance of the K\"ahler potential part
of ${\cal L}$ corresponds then to the condition $w K_\Phi \Phi = K$ and
\beq
\label{WZ7c}
\Delta = 2(K- w K_\Phi \Phi)
\eeq
measures the violation of scale invariance. Similarly, the holomorphic quantity
\beq
\label{WZ7d}
\widetilde\Delta = 3W - wW_\Phi\Phi
\eeq
measures the violation of scale invariance in the superpotential terms.
Actually, if $j_\mu^{(dil.)}$ is the Noether current for dilatations, we have on-shell\footnote{We use the expansion $\Phi = z + \sqrt2\theta\psi - \theta\theta f$. Furthermore, since 
$K_\Phi\Phi = \ov\Phi K_{\ov\Phi}$, $zK_{zz\ov z} = \ov z K_{\ov z\ov z z}$.}
\beq
\label{WZ7e}
\begin{array}{rcl}
\partial^\mu j_\mu^{(dil.)} &=& 2(w-1)K_{z\ov z} [ (\partial^\mu\ov z)(\partial_\mu z) + \ov f f ]
- (w-3)[ W_zf + \ov f\ov W_{\ov z}]
\crbig
&& + 2wK_{\ov z z z} z [ (\partial^\mu\ov z)(\partial_\mu z) + \ov f f ]
-wW_{zz}zf - w \ov W_{\ov{zz}}\ov z \ov f
\crbig
&&\makebox{$+$ fermions $+$ gauge terms }
\crbig
&=& - \Delta |_{\theta\theta\ov\theta\ov\theta} - \widetilde\Delta|_{\theta\theta}
- \ov{\widetilde\Delta}|_{\ov{\theta\theta}} 
- {1\over4}\Box \Delta|_{\theta=0} + \makebox{gauge terms }
\end{array}
\eeq
and the gauge terms are proportional to $wg_z z$ or $w\ov z \ov g_{\ov z}$. 
Hence scale invariance holds
if 
\beq
\label{WZ7f}
\Delta = \widetilde \Delta = wg_z z = 0.
\eeq
The Lagrangian induces field equations
\beq
\label{WZ8}
\overline{DD}K_\Phi = 4\,W_\Phi + g_\Phi \widetilde\Tr({\cal W}^\alpha {\cal W}_\alpha)
\eeq
for the chiral superfield $\Phi$ and, for $A$, 
\beq
\label{WZ9}
(g+\ov g) D^\alpha (e^A{\cal W}_\alpha e^{-A}) = 2 \,T( R ) K^\prime e^A\Phi\ov\Phi
- (D^\alpha g) e^A {\cal W}_\alpha e^{-A}
- (\ov D_\dalpha\ov g) \ov {\cal W}^\dalpha \, .
\eeq
The Bianchi identity (\ref{WZ10}) has been used to simplify the field equations for $A$.

Next, we insert the field equations into our identities. The resulting equations then hold
only for on-shell fields. 
Using the first identity (\ref{WZ3}) and matter field equation (\ref{WZ8}), one obtains
\beq
\label{WZ11}
-2 \, \ov D^\dalpha\Bigl[ (\ov{\cal D}_\dalpha\ov\Phi) K_{\Phi\ov\Phi}({\cal D}_\alpha\Phi) \Bigr] =
\ov{DD}D_\alpha K + 4 D_\alpha W
+4 K_\Phi {\cal W}_\alpha \Phi + \widetilde\Tr({\cal W}^\beta {\cal W}_\beta) D_\alpha g.
\eeq
Using the second identity (\ref{WZ10c}) and the gauge field equation (\ref{WZ9}), one also gets
\beq
\label{WZ12}
-2\,\ov D^\dalpha\Bigl[ (g+\ov g) \widetilde\Tr ({\cal W}_\alpha e^{-A} \ov{\cal W}_\dalpha e^A) \Bigr] = 
-4 K_\Phi {\cal W}_\alpha \Phi - \widetilde\Tr({\cal W}^\beta {\cal W}_\beta) D_\alpha g.
\eeq
Comparing, one finds the following supercurrent structure:
\beq
\label{WZ14}
\begin{array}{rcl}
\ov D^\dalpha\, J_{\alpha\dalpha} &=& D_\alpha X + \chi_\alpha , 
\crbig
J_{\alpha\dalpha} &=& -2 (\ov{\cal D}_{\dalpha}\ov\Phi)K_{\Phi\ov\Phi} ({\cal D}_\alpha\Phi)
- 2 (g+\ov g)\widetilde\Tr ({\cal W}_\alpha e^{-A}\ov {\cal W}_\dalpha e^A) ,
\crbig
X &=& 4W ,
\crbig
\chi_\alpha &=& \ov{DD}D_\alpha K.
\end{array}
\eeq
One may view these equations as the {\it natural} supercurrent structure for the gauged 
Wess-Zumino model. It uses the full $16_B+16_F$ structure.\footnote{This corresponds to the 
${\cal S}$--multiplet of ref.~\cite{KS}.}
All quantities are gauge-invariant and also invariant under K\"ahler transformations 
$K \rightarrow K + \Xi(\Phi) + \ov \Xi(\ov\Phi)$. And, as we discuss below,
the energy-momentum tensor included in $J_{\alpha\dalpha}$ is the canonical (Noether) tensor,
improved to the symmetric gauge-invariant (Belinfante) tensor\footnote{Strictly speaking, the
difference between the Belinfante and the canonical (Noether) energy-momentum tensors is not
an improvement. It uses the field equations for gauge fields.}. The improvement transformation (\ref{impr2}) can in principle be used to reduce the structure, but a nonzero superpotential 
is not in general of the form $-\ov{DD}\,{\cal G}$ and $X$ cannot then be removed. 
There are exceptions. For instance, if the superpotential is homogeneous,  
$W_\Phi\Phi = \ell W$, and if $g_\Phi\Phi=0$, the field equation implies $X = 4W = \ell^{-1}\ov{DD}K_\Phi\Phi$.
Canceling $\chi_\alpha$ is always possible, at the price however of losing K\"ahler invariance
of the transformed $J_{\alpha\dalpha}$ and $X$, with consequences explained in ref.~\cite{KS}.
Notice also that while the chiral anomaly term $D_\alpha X$ is generated by the field equation for $\Phi$, 
the linear anomaly $\chi_\alpha$ is truly off-shell: it is already present in the identity (\ref{WZ3}).

Theory (\ref{WZ7}) has kinetic metrics ${1\over2}(g+\ov g)\delta_{ab}$ for the gauge multiplet
and $K_{z\ov z}$ for the chiral superfield 
components.
The lowest component of $J_{\alpha\dalpha}$ is the fermionic current\footnote{We use the notation $\lambda$ for gauginos.}
\beq
\label{WZ15}
\begin{array}{rcl}
j_\mu \,\,\equiv\,\, {3\over8}(\ov\sigma_\mu)^{\dalpha\alpha} J_{\alpha\dalpha} |_{\theta=0} &=&
 {3\over2} \, \ov\psi K_{z\ov z} \ov\sigma_\mu\psi - {3\over4}(g+\ov g) \, \ov\lambda\ov\sigma_\mu\lambda
\crbig
&=& {3\over4} \Bigl[ \ov\psi K_{z\ov z}\,\gamma_\mu\gamma_5\psi 
- {1\over2}(g+\ov g) \ov\lambda\gamma_\mu\gamma_5\lambda \Bigr].
\end{array}
\eeq
This is the current of chiral $U(1)$ rotations of the (two-component) fermion fields with charges 
$3/2$ for the
gaugino (we choose this normalization for $R$--transformations of gauginos and Grassmann
superspace coordinates) and $-3/2$ for chiral fermions. Hence, $j_\mu$ is the current of the
$U(1)_{\widetilde R}$ group leaving chiral superfields inert, which is a symmetry for all K\"ahler potentials
$K$ and gauge kinetic functions $g$ if the superpotential (and then $X$) vanishes.\footnote{
For a generic $R$--symmetry, transformations are
$$
\begin{array}{rcl}
\theta \quad&\longrightarrow&\quad e^{3i\alpha/2}\,\theta,
\\
{\cal W}_\alpha \quad&\longrightarrow&\quad e^{3i\alpha/2} \,{\cal W}_\alpha , 
\\
\Phi \quad&\longrightarrow&\quad e^{iq\alpha}\,\Phi.
 \end{array}
$$
The case $q=0$ corresponds to $U(1)_{\widetilde R}$ transformations.
}
Since however $\chi_\alpha$ is never zero, the energy-momentum tensor included in 
$J_{\alpha\dalpha}$ is never traceless, even if the theory is scale-invariant.

In the expansion (\ref{4sc8}) of the supercurrent superfield (\ref{WZ14}) we find the following 
bosonic energy-momentum tensor:
\beq
\label{WZ16}
\begin{array}{rcl}
T_{\mu\nu} &=& (D_\mu \ov z) K_{z\ov z} (D_\nu z) 
+ (D_\nu\ov z) K_{z\ov z} (D_\mu z)
-{1\over2} (g+\ov g) F_{\mu\rho}^a {F_\nu^a}^\rho
\crbig
&& -\eta_{\mu\nu} \Bigl[ (D_\rho\ov z) K_{z\ov z} (D^\rho z) 
-{1\over8} (g+\ov g) F_{\rho\sigma}^a F^{a\,\rho\sigma} + \ov f K_{z\ov z} f -{1\over 4} (g+\ov g)d^a d^a\Bigr]
\crbig
&& + {1\over2} \,\eta_{\mu\nu} \Re f_X,
\end{array}
\eeq
where
\beq
\begin{array}{rcl}
D_\mu z &=& \partial_\mu z+\frac{i}{2}T^a z A^a_\mu , 
\crbig
F^a_{\mu\nu} &=& \partial_\mu A^a_\nu-\partial_\nu A^a_\mu-\frac{1}{2}f^{abc}A^b_\mu A^c_\nu
\end{array}
\eeq
are the gauge-covariant derivatives of the scalars and the gauge-field strength tensor
respectively. Since $X=4W$, $\Re f_X= 2(W_z f + \ov f\ov W_{\ov z})$. As already stated,
besides auxiliary field contributions which we discuss below, 
$T_{\mu\nu}$ is the canonical (Noether) tensor, improved to the symmetric gauge-invariant 
(Belinfante) tensor. 

The auxiliary field structure is interesting. The bosonic contributions involving $f$ 
(in $\Phi = -\theta\theta f + \ldots$) and $d^a$ (in $A^a = {1\over2}\theta\theta\ov{\theta\theta} d^a
+ \ldots$) are:
\beq
\label{WZ17}
\begin{array}{lrcl}
\makebox{In $T_{\mu\nu}$:}&\qquad T_{\mu\nu}^{(aux.)} &=& 
- \eta_{\mu\nu} \Bigl[  \ov f K_{z\ov z} f  - W_zf - \ov f \ov W_{\ov z}
- {1\over4}(g+\ov g) d^ad^a \Bigr] .
\crbig
\makebox{In $X$:}&\qquad f_X^{(aux.)} &=& 4 W_z\, f .
\crbig
\makebox{In $\chi_\alpha$:}&\qquad D^{(aux.)} &=& - 4 K_z T^a z  \, d^a - 8 \ov f K_{z\ov z} f.
\crbig
\makebox{In eqs.~(\ref{4sc10}):}&\qquad {T^\mu}_\mu ^{(aux.)} &=& 
- K_z T^a z \, d^a - 2\ov f K_{z\ov z} f + 3 (W_z\, f + \ov f \ov W_{\ov z}) ,
\crbig
&  \partial^\mu j_\mu ^{(aux.)} &=& -6\Im (W_z\,f) .
\end{array}
\eeq
If the energy-momentum tensor were expressed as a function of off-shell auxiliary 
fields, the Noether procedure would lead to
\beq
\label{WZ17b}
T_{\mu\nu}^{(aux.)} = - \eta_{\mu\nu} {\cal L}^{(aux.)} = 
- \eta_{\mu\nu} \Bigl[  \ov f K_{z\ov z} f  - W_zf - \ov f \ov W_{\ov z}
+ {1\over4}(g+\ov g) d^ad^a + {1\over2} d^a K_z T^a z  \Bigr] .
\eeq
It turns out that the component expansion (\ref{4sc8}) provides the off-shell expression for
chiral superfields, but not for gauge fields. One can check that if $X\ne0\ne\chi_\alpha$, the
auxiliary field energy-momentum tensor included in $J_{\alpha\dalpha}$ is never completely ``off-shell",
essentially because the component expansion of $J_{\alpha\dalpha}$ follows from the supercurrent equation which holds on-shell.
Using the first eq.~(\ref{4sc10}) does not help.
Instead, replacing auxiliary fields by their on-shell values, 
\beq
K_{z\ov z}f = \ov W_{\ov z}, \qquad\qquad
(g+\ov g) d^a = - K_z T^a z,
\eeq
we find then
$T_{\mu\nu}^{(aux.)} = \eta_{\mu\nu} {\cal V}$ with scalar potential
\beq
{\cal V} = \ov f K_{z\ov z} f + {g+\ov g\over4} d^ad^a
= W_z(K_{z\ov z})^{-1} \ov W_{\ov z} + {1\over 4(g+\ov g)} \sum_a (K_z T^a z)^2,
\eeq
and all equations in (\ref{WZ17}) are of course consistent with the anomaly relations
(\ref{4sc10}). Hence, the correct identification of
the energy-momentum tensor, even if $J_{\alpha\dalpha}$ is expressed in terms of 
off-shell superfields, holds on-shell only. 
Since
\beq
\label{WZ18}
\widetilde F^a_{\mu\rho} {F_\nu^a}^\rho 
- {1\over 4}\eta_{\mu\nu} \, \widetilde F_{\rho\sigma}^a F^{a\rho\sigma}
\equiv 0 ,
\eeq
($\widetilde F^a_{\mu\nu}={1\over 2}\epsilon_{\mu\nu\rho\sigma}F^{a\rho\sigma}$)
the energy-momentum tensor in $J_{\alpha\dalpha}$ [eq.~(\ref{WZ16})] does not depend on
$\widetilde F^a_{\mu\nu}$ and then on $\Im g(z)$.

Theory (\ref{WZ7}) is expected to be scale-invariant if $\Delta=\widetilde\Delta
= wg_zz=0$ [eq.~(\ref{WZ7f})]. 
Since however $\chi_\alpha\neq 0$ in the supercurrent structure (\ref{WZ14}), 
the energy-momentum tensor (\ref{WZ16}) is never traceless, even if the theory is 
scale-invariant. In the canonical formulation leading to energy-momentum tensor (\ref{WZ16})
the dilatation current $j_\mu^{(dil.)}$ verifies
\beq
\label{WZ19}
j_\mu^{(dil.)}=
x^\nu T_{\mu\nu}^{(can.)}+ \sum_i w_i\frac{\partial\mathcal{L}}{\partial\partial^\mu\varphi_i}\varphi_i,
\eeq
where $\varphi_i$ is a generic field with scale dimension $w_i$ and 
$T_{\mu\nu}^{(can.)} = \sum_i {\partial{\cal L}\over\partial\partial^\mu\varphi_i}\partial_\nu\varphi_i
- \eta_{\mu\nu}{\cal L}$ is the canonical energy-momentum tensor. Hence,
\beq
\label{WZ20}
T^{(can.)\mu}{}_\mu=\partial^\mu j_\mu^{(dil.)}
- \sum_i w_i \, \partial^\mu\left(\frac{\partial\mathcal{L}}{\partial\partial^\mu\varphi_i}\varphi_i\right),
\eeq
which is nonzero if $\partial^\mu j_\mu^{(dil.)}=0$. 

Callan, Coleman and Jackiw (CCJ) \cite{CCJ}\footnote{See for instance ref.
\cite{Ortin}, section 2.4, for a general discussion.} have demonstrated how to improve the 
energy-momentum tensor into a new expression $\Theta_{\mu\nu} = \Theta_{\nu\mu}$ which verifies
\beq
\label{WZ21}
 j_\mu^{(dil.)}=x^\nu \Theta_{\mu\nu} \, , \qquad\qquad
 \partial^\mu j_\mu^{(dil.)}= {\Theta^\mu}_\mu \, ,
\eeq
and is on-shell traceless if the theory is scale-invariant. 
We now want to discuss two different supersymmetric improvements of the ``natural" 
supercurrent structure (\ref{WZ14}) leading either to a Ferrara-Zumino structure with 
$\chi_\alpha=0$ or to the CCJ energy-momentum tensor. 

\subsection{Improvement to the Ferrara-Zumino supercurrent }

The supersymmetric improvement transformation (\ref{impr2}) is generated by a real 
superfield ${\cal G}$. We first choose 
\beq
\label{WZ25}
{\cal G}=-\frac{1}{3}K
\eeq
to eliminate $\chi_\alpha$ in the supercurrent structure (\ref{WZ14}). The result is
the commonly used $12_B+12_F$ Ferrara-Zumino structure
\beq
\label{WZ26}
\begin{array}{rcl}
\ov D^\dalpha\, J^{(1)}_{\alpha\dalpha} &=& D_\alpha X^{(1)}, 
\crbig
J^{(1)}_{\alpha\dalpha} &=& -2 (\ov{\cal D}_{\dalpha}\ov\Phi)K_{\Phi\ov\Phi} ({\cal D}_\alpha\Phi)
- 2 (g+\ov g)\widetilde\Tr ({\cal W}_\alpha e^{-A}\ov {\cal W}_\dalpha e^A)-{2\over 3}\,[ D_\alpha , \ov D_\dalpha ] \,K ,
\crbig
X^{(1)} &=& 4W-{1\over 3}\ov{DD}K .
\end{array}
\eeq
Since $X^{(1)}$ and $J^{(1)}_{\alpha\dalpha}$ are not invariant under K\"ahler transformations, an obstruction to this  
transformation could arise if the global description of the sigma-model target space requires 
patches with K\"ahler potentials linked by K\"ahler transformations \cite{KS}.

Using field equation (\ref{WZ8}), the chiral anomaly superfield $X^{(1)}$ can be written as
\beq
\label{WZ27}
 X^{(1)}=\frac{4}{3}\widetilde \Delta-\frac{1}{6}\ov{DD}\Delta-\frac{w}{3}g_\Phi\Phi\widetilde \Tr({\cal W}^\alpha {\cal W}_\alpha).
\eeq
It vanishes in a scale-invariant theory where $\Delta=\widetilde\Delta = wg_\Phi\Phi=0$. Since $\chi_\alpha^{(1)}=0$,
we have
\beq
\label{WZ27b}
{T^{(1)\mu}}_\mu = {3\over2}\Re f_{X^{(1)}}
\eeq
and the energy-momentum tensor in $J^{(1)}_{\alpha\dalpha}$ is traceless in a scale-invariant theory
where $\partial^\mu j_\mu^{(dil.)} = 0$. However, for a generic theory without scale invariance, 
we have
\beq
{T^{(1)\mu}}_\mu = {3\over2}\Re f_{X^{(1)}}
\ne
\partial^\mu j_\mu^{(dil.)} \ne 0.
\eeq
The difference between the ``natural" energy-momentum tensor $T_{\mu\nu}$ [eq.~(\ref{WZ16})] and the 
improved version is
\beq
T_{\mu\nu}^{(1)} - T_{\mu\nu} =  -{1\over3}(\partial_\mu\partial_\nu - \eta_{\mu\nu}\Box)K ,
\eeq
which is not the improvement to the  tensor $\Theta_{\mu\nu}$ found by CCJ.

For clarity of the next equations, we now use the notation
\beq
K_i=K_{z^i} = {\partial K\over\partial z^i},\quad\quad 
K^i=K_{\ov z_i} = {\partial K\over\partial \ov z_i}, \qquad\qquad \ldots
\eeq
The lowest component of the supercurrent superfield $J^{(1)}_{\alpha\dalpha}$ then reads
\beq
\label{WZ28}
j^{(1)}_\mu = {1\over 2}\sum_{i,j}K^j_i\ov\psi_j\ov\sigma_\mu\psi^i+i\sum_i\left(K^i D_\mu \ov z_i-K_i D_\mu z^i\right)-{3\over 4}(g+\ov g)\ov\lambda\ov\sigma_\mu\lambda.
\eeq
This current has the following origin. Suppose that we start with a superconformal gauge theory, as in the superconformal construction
of the $N=1$ supergravity--matter system \cite{CFGVP}. We have then
an auxiliary vector field $A_\mu$ to gauge the $U(1)_R$ symmetry inside 
$SU(2,2|1) \supset SU(2,2)\times U(1)_R$. The field equation for this
auxiliary field is then\footnote{See ref.~\cite{KU}, eq.~(33).}
\beq
A_\mu \sim j_\mu^{(1)}.
\eeq
Even if the fermionic terms seem correct, the current $j_\mu^{(1)}$ is not the Noether current of the
$R$--symmetry with canonical weights 1 for chiral superfields and 3/2 for gauginos. For the chiral 
superfield terms, it should be viewed as the K\"ahler connection. 
This chiral current is not in general conserved, except if $f_{X^{(1)}}=0$, {\it i.e.} if the theory has 
conformal symmetry. 

Hence, this improved supercurrent structure with $12_B+12_F$ fields does not include particularly 
attractive currents, in obvious relation with the potential symmetries of the 
globally supersymmetric gauge theory.
But it may be useful in the coupling to (linearized) supergravity, using the
old minimal $12_B+12_F$ supergravity off-shell multiplet.

\subsection{Improvement to the CCJ supercurrent structure}

Suppose instead that we choose
\beq
\label{WZ30}
{\cal G} = -{1\over6}\, w ( K_\Phi \Phi + \ov\Phi K_{\ov\Phi} )
= -{1\over3} \, w K_\Phi \Phi \, .
\eeq
The number $w$ will be, as earlier, identified with the scale weight of all chiral 
superfields. The resulting supercurrent structure is
\beq
\label{WZ31}
\begin{array}{rcl}
\ov D^\dalpha\, J^{(2)}_{\alpha\dalpha} &=& D_\alpha X^{(2)} + \chi^{(2)}_\alpha , 
\crbig
J^{(2)}_{\alpha\dalpha} &=& -2 (\ov{\cal D}_{\dalpha}\ov\Phi)K_{\Phi\ov\Phi} ({\cal D}_\alpha\Phi)
- 2 (g+\ov g)\widetilde\Tr ({\cal W}_\alpha e^{-A}\ov {\cal W}_\dalpha e^A)
- {2\over 3}w\,[ D_\alpha , \ov D_\dalpha ] \,K_\Phi \Phi ,
\crbig
X^{(2)} &=& 4W-{w\over 3}(\ov{DD}K_\Phi)\Phi ,
\crbig
\chi^{(2)}_\alpha &=& \ov{DD}D_\alpha(K-wK_\Phi \Phi)
\,\,=\,\, {1\over2} \ov{DD}D_\alpha \Delta.
\end{array}
\eeq
The improved supercurrent superfield $J_{\alpha\dalpha}^{(2)}$ is not K\"ahler-invariant.

The first reason to consider this particular structure is that the lowest component of $J_{\alpha\dalpha}$ reads
\beq
\label{WZ32}
\begin{array}{rcl}
j^{(2)}_\mu &=& \displaystyle iw\sum_{i,j} K^j_i\left(z^i D_\mu \ov z_j-\ov z_j D_\mu z^i\right)
- {1\over 2}w\sum_{i,j,k} \ov\psi_j\ov\sigma_\mu\psi^i\left(z^k K^j_{ki}+\ov z_k K^{kj}_i\right)
\crbig
&& \displaystyle - \left(w-{3\over 2}\right)\sum_{i,j}K^j_i\ov\psi_j\ov\sigma_\mu\psi^i 
- {3\over 4}(g+\ov g)\ov\lambda\ov\sigma_\mu\lambda.
\end{array}
\eeq
This is precisely the Noether current of the $R$ transformation, under which chiral superfields 
have $R$--charge $q=w$. On-shell, the superfield $X^{(2)}$ reads
\beq
\label{WZ33}
X^{(2)} = {4\over3}\,\widetilde\Delta
- {w\over 3}g_\Phi\Phi\widetilde\Tr({\cal W}^\alpha {\cal W}_\alpha).
\eeq
Since we assume that $K_\Phi\Phi = \ov\Phi K_{\ov\Phi}$, the K\"ahler part of the theory is always 
$R$--invariant, for all $q$.
Hence, $X^{(2)}$, which measures $\partial^\mu j_\mu^{(2)}$ should not, as in eq.~(\ref{WZ27}), depend on $K$.
$R$--symmetry follows if $\widetilde\Delta=wg_\Phi\Phi =0$, {\it i.e.} if $X^{(2)}$ vanishes. And
as usual, scale invariance requires $\Delta=\widetilde\Delta=wg_\Phi\Phi=0$, implying $R$--invariance.

The second reason to consider this supercurrent structure is that with this improvement,
the superfield $J_{\alpha\dalpha}^{(2)}$ contains the CCJ energy-momentum tensor. 
One can actually verify that
\beq
\label{WZ34b}
{T^{(2)\mu}}_\mu = {3\over2}\Re f_{X^{(2)}} + {1\over4} D^{(2)} = \partial^\mu j_\mu^{(dil.)} 
\eeq
in general, even if the theory is not scale-invariant.

Hence, this second improvement leads to a supercurrent structure with energy-momentum tensor
$$
T^{(2)}_{\mu\nu} = \Theta_{\mu\nu}, \qquad\qquad
{\Theta^\mu}_\mu = \partial^\mu j_\mu^{(dil.)},
$$
and the Noether current of the $R$ transformation under which chiral superfields have charge $q=w$.
 
Finally, if $\Delta=0$ (and then $\chi^{(2)}_\alpha=0$), {\it i.e.} if $K=wK_\Phi\Phi$, the K\"ahler potential 
part of the theory is scale-invariant, and both improvements are of course identical. But they significantly 
differ in a generic theory.

\section{The Fayet-Iliopoulos term}\label{secFIterms}
\setcounter{equation}{0}

We now consider the case of super-Maxwell theory with a Fayet-Iliopoulos term, where 
a (non--$R$) $U(1)$ symmetry is gauged. This
model has been considered by several authors\footnote{See for instance \cite{Kuz2, KS2, DT}
for recent discussions.}, but we wish to present
a somewhat different and unorthodox approach. 
The theory includes chiral superfields $\phi_i$
with $U(1)$ charges $q_i$ (we label both $\phi_i$ and $\ov\phi_i$ with the same lower index here) and gauge transformations
\beq
\label{FI1}
\phi_i \qquad\longrightarrow\qquad e^{q_i\Lambda}\phi_i, \qquad\qquad \ov D_\dalpha\Lambda=0
\eeq
and an abelian gauge superfield $A$ with $\delta A=-\Lambda-\ov\Lambda$.
Gauge-covariant supersymmetric derivatives are
\beq
\label{FI2}
\begin{array}{rcccl}
{\cal D}_\alpha\phi_i &=& e^{-q_i A}D_\alpha e^{q_iA} \phi_i &=& D_\alpha \phi_i + q_i(D_\alpha A)\phi_i,
\crbig
\ov{\cal D}_\dalpha\ov\phi_i &=& e^{-q_i A}\ov D_\dalpha e^{q_iA} \ov\phi_i &=& 
\ov D_\dalpha\ov\phi_i
+ q_i(\ov D_\dalpha A) \ov\phi_i 
\end{array}
\eeq
and identity (\ref{WZ3}) becomes
\beq
\label{FI3}
2 \, \ov D^\dalpha\sum_{i,j}\Bigl[ K_{\phi_i\ov\phi_j} ({\cal D}_\alpha\phi_i)(\ov{\cal D}_\dalpha\ov\phi_j) \Bigr] =
\ov{DD}D_\alpha K + 4 \sum_i q_iK_{\phi_i}\phi_i {\cal W}_\alpha
+ \sum_i(\ov{DD}K_{\phi_i})  ({\cal D}_\alpha\phi_i) .
\eeq
Using Lagrangian
\beq
\label{FI4}
{\cal L} = \Dint \Bigl[ K(\bar\phi_i e^{q_iA}\phi_i)+\xi A \Bigr]
+\Fint \left[ W(\phi_i)+\frac{1}{4}g(\phi_i){\cal W}^\alpha {\cal W}_\alpha\right]+  {\rm h.c.},
\eeq
where $\xi$ is the (dimension-two) Fayet-Iliopoulos (FI) coefficient and
\beq
\label{FI5}
{\cal W}_\alpha = -{1\over4}\ov{DD}D_\alpha A, \qquad\qquad
\ov {\cal W}_\dalpha = -{1\over4} DD \ov D_\dalpha A,
\eeq
the field equations are
\beq
\label{FI6}
\begin{array}{rcl}
\ov{DD}\, K_{\phi_i} &=& 4 W_{\phi_i} + g_{\phi_i}\, {\cal W}^\alpha {\cal W}_\alpha ,
\crbig
\ov D_\dalpha \Bigl[ (g+\ov g)\ov {\cal W}^\dalpha \Bigr] &=& 2\xi + 2\sum_i q_iX_i {\partial K\over\partial X_i}
- (D^\alpha g){\cal W}_\alpha,
\qquad\qquad
X_i = \ov\phi_i e^{q_iA} \phi_i .
\end{array}
\eeq
The abelian Bianchi identity $D^\alpha {\cal W}_\alpha = \ov D_\dalpha\ov {\cal W}^\dalpha$ has 
been used. The two functions $W(\phi_i)$ and $g(\phi_i)$ are $U(1)$--invariant: 
$\sum_i q_i\phi_i W_{\phi_i} = \sum_iq_i\phi_i g_{\phi_i} =0$. Then, the first field equation also implies
$\ov{DD}\sum_i q_i  \phi_i K_{\phi_i} = 0$.

The real Lagrangian superfield $K+\xi A$ is not invariant under 
supersymmetric gauge transformations.\footnote{Or K\"ahler transformations.} But the
non-invariance of the Lagrangian is confined to a derivative which does not generate dynamics 
and vanishes in the Wess-Zumino gauge. Using the expansion
\beq
\label{FI6b}
A = C + \theta\sigma^\mu\ov\theta\, A_\mu + \theta\theta\ov{\theta\theta}\Bigl[ {1\over2} d
- {1\over4}\Box C \Bigr] + \ldots ,
\eeq
$C$ transforms and vanishes in Wess-Zumino gauge, 
$d$ is gauge-invariant and we may use 
$$
{\cal L}_{FI} = \xi \Bigl[ {1\over2} d - {1\over4}\Box C \Bigr]
$$
as the FI Lagrangian.
Field equations (\ref{FI6}) are then invariant or covariant under supersymmetric gauge transformations. Hence, the absence of gauge invariance of the Lagrangian superfield is 
irrelevant to the Lagrangian and to the dynamical equations. 
However, in the canonical formalism, a derivative contribution does induce, in particular, an
energy-momentum tensor which does not affect the total energy-momentum of the theory, 
which is the physically significant quantity.\footnote{In other words, it adds 
an improvement term to the physically relevant energy-momentum tensor.}

In the supercurrent structure, one insists in working with superfields and then one will
plausibly face gauge non-invariance of operators which are however not physical quantities. 
This is not so for the natural supercurrent structure. 
As in the non-abelian case, field equations imply the gauge-invariant equations
\beq
\label{FI7}
\begin{array}{rcl}
\displaystyle \sum_i (\ov{DD}\,K_{\phi_i} )({\cal D}_\alpha\phi_i) &=& 4\,D_\alpha W + (D_\alpha g) {\cal W}^\beta {\cal W}_\beta ,
\crbig
\ov D^\dalpha \Bigl[ (g+\ov g) {\cal W}_\alpha \ov {\cal W}_\dalpha \Bigr]
&=& \displaystyle 2\,\xi\, {\cal W}_\alpha + 2 \sum_i q_iX_i{\partial K\over\partial X_i} {\cal W}_\alpha
+ {1\over2} (D_\alpha g) {\cal W}^\beta {\cal W}_\beta.
\end{array}
\eeq
For fields solving field equations, identity (\ref{FI3}) and the abelian version of formula (\ref{WZ10c}) 
turn then into the ``natural" $16_B+16_F$ 
supercurrent structure
\beq
\label{FI8}
\begin{array}{rcl}
\ov D^\dalpha J_{\alpha\dalpha} &=& D_\alpha X + \chi_\alpha , 
\crbig
J_{\alpha\dalpha} &=& \displaystyle 2 \,\sum_{i,j} K_{\phi_i\ov\phi_j} ({\cal D}_\alpha\phi_i)(\ov{\cal D}_\dalpha\ov\phi_j)
- 2(g+\ov g){\cal W}_\alpha\ov {\cal W}_\dalpha ,
\crbig
X &=& 4W, 
\crbig
\chi_\alpha &=& \ov{DD}D_\alpha K - 4\,\xi\,{\cal W}_\alpha \,\,=\,\, \ov{DD}D_\alpha (K+\xi A).
\end{array}
\eeq
The Fayet-Iliopoulos coefficient appears, as it should, as a source of scale invariance breaking in
$\chi_\alpha$. Its contribution is as expected gauge-invariant.\footnote{For super-Maxwell theory 
($\phi_i=0$), this is the supercurrent structure obtained by Kuzenko \cite{Kuz2}.}
The energy-momentum tensor included in $J_{\alpha\dalpha}$ depends 
on $\xi$, according to the first condition (\ref{4sc10}), but only via the on-shell
value of the auxiliary field $d$.\footnote{See the discussion following
eqs.~(\ref{WZ17}).}

\subsection{The improved supercurrent structure}

To obtain the supercurrent superfield which contains the new improved energy-mo\-men\-tum tensor 
$\Theta_{\mu\nu}$ and the appropriate $U(1)_R$ current, we now transform the supercurrent structure (\ref{FI8}) with the real superfield $\mathcal{G}=-w \sum_i\phi_i K_{\phi_i}/3$, $w$ being as usual the (common) scale dimension of the chiral superfields. This leads to the supercurrent structure
\beq
\label{FI9}
\begin{array}{rcl}
\widetilde J_{\alpha\dot\alpha} &=& \displaystyle 2\,\sum_{i,j}K_{\phi_i\bar\phi_j}(\mathcal{D}_\alpha\phi_i)(\ov{\mathcal{D}}_{\dot\alpha}
\bar\phi_j)-\frac{2w}{3}[D_\alpha,\ov{D}_{\dot\alpha}]\sum_i\phi_i K_{\phi_i}
- 2(g+\ov g){\cal W}_\alpha\ov {\cal W}_{\dot\alpha} ,
\crbig
\widetilde X &=& \displaystyle 4W-\frac{w}{3}\sum_i\phi_i\ov{DD}K_{\phi_i} ,
\crbig
\widetilde\chi_\alpha &=& {1\over 2}\ov{DD}D_\alpha\Delta-4\xi {\cal W}_\alpha.
\end{array}
\eeq
As in the non-abelian case, using the first of field equations (\ref{FI6}), 
$\widetilde X$ can be written as
\beq
\label{FI10}
\widetilde X={4\over 3}\widetilde\Delta-\frac{w}{3}\sum_i\phi_ig_{\phi_i}{\cal W}^\alpha {\cal W}_\alpha.
\eeq
and the K\"ahler potential only contributes to scale symmetry violation in $\chi_\alpha$.
The condition $\widetilde\Delta=0$ on the superpotential is necessary for
$U(1)_R$ invariance of the theory. If in addition $\sum_i\phi_ig_{\phi_i}=0$, then $\widetilde X=0$ and the lowest component of $\widetilde J_{\alpha\dot\alpha}$ is conserved, as it should. Finally, the dimensionful
Fayet-Iliopoulos term always breaks scale invariance without affecting the $R$-symmetry.

\subsection{On the Ferrara-Zumino supercurrent structure with a \\
Fayet-Iliopoulos term}

Our goal is to remove, using an improvement transformation (\ref{impr2}), 
the contribution $\chi_\alpha = -4\,\xi {\cal W}_\alpha = \xi\,\ov{DD}D_\alpha A$.
But we want a gauge-invariant procedure: we cannot use ${\cal G} = -\xi A/3$ in improvement 
transformation (\ref{impr2}). But, as we observed earlier, the problem with gauge invariance
is related to our prejudice to work with superfields and to the related 
presence of 
a derivative term in the FI Lagrangian. We then first make this derivative term gauge-invariant:
we formally introduce a new chiral superfield $S$ and postulate
that the quantity $A+S+\ov S$ is invariant under supersymmetric gauge transformations. Since 
$$
\ov{DD} D_\alpha (A+S+\ov S) = -4\,{\cal W}_\alpha,
$$
we prefer to consider the theory 
\beq
\label{FI11}
{\cal L} = \Dint \Bigl[ K(\bar\phi_i e^{q_iA}\phi_i)+\xi (A + S + \ov S )\Bigr]
+\Fint \left[ W(\phi_i)+\frac{1}{4}g(\phi_i){\cal W}^\alpha {\cal W}_\alpha\right]+  {\rm h.c.},
\eeq
with a gauge-invariant Lagrangian {\it superfield}. With expansion
\beq
\label{FI12}
A+S+\ov S = C + 2\Re s + \ldots + {1\over2} \theta\theta\ov{\theta\theta} 
\Bigl[ d - {1\over2}\Box (C + 2\Re s) \Bigr],
\eeq
we find
\beq
\label{FI13}
\xi \Dint \Bigl[ A + S + \ov S \Bigr] = \xi \Bigl[ {1\over2} d - {1\over4}\Box
(C + 2\Re s) \Bigr].
\eeq
Since $S$ does not generate any dynamics\footnote{This is true in all gauges.}, we are considering
the same Fayet-Iliopoulos term as in theory (\ref{FI4}), but defined now in terms of 
a real gauge-invariant superfield $A+S+\ov S$. Notice that the derivative terms in the Lagrangian will
generate the following contribution to the canonical energy-momentum tensor:
\beq
\label{FI13a}
T^{(S)}_{\mu\nu} = - {\xi\over4}\, 
(\partial_\mu\partial_\nu - \eta_{\mu\nu} \Box) (C + 2\Re s) ,
\qquad\quad
{T^{(S)\mu}}_\mu = {3\over4} \,\xi\, \Box (C + 2\Re s).
\eeq
But this is an improvement term that will not affect the Noether charges. Since the natural supercurrent
structure (\ref{FI8}) is unaffected by the introduction of $S$, it does not contain the improvement term
(\ref{FI13a}). Notice also that a K\"ahler transformation generates an
improvement contribution similar to eq.~(\ref{FI13a}). Hence it can be used 
to eliminate the field $S$, confirming the formal character of its introduction. 

We may then use the gauge-invariant superfield
\beq
\label{FI14}
{\cal G} = -{1\over3} [K + \xi(A+S+\ov S)]
\eeq
in improvement transformation (\ref{impr2}), to obtain the Ferrara-Zumino supercurrent structure
\beq
\label{FI15}
\begin{array}{rcl}
\widehat J_{\alpha\dalpha} &=& \displaystyle 2\sum_{i,j} K_{\phi_i\ov\phi_j} ({\cal D}_\alpha\phi_i)(\ov{\cal D}_\dalpha\ov\phi_j)
- 2(g+\ov g){\cal W}_\alpha\ov {\cal W}_\dalpha 
\crbig
&& \hspace{1.9cm}
\displaystyle
- {2\over 3}[D_\alpha,\ov{D}_\dalpha]\Bigl[K+\xi(A+S+\bar S)\Bigr],
\crbig
\widehat X &=& \displaystyle 4W-{1\over 3}\ov{DD}[K+\xi(A+S+\bar S)]
\end{array}
\eeq
with $\widehat \chi_\alpha = 0$. All quantities are gauge-invariant.
The lowest component of the supercurrent superfield (\ref{FI15}) reads
\beq
\label{FI16}
\widehat j_\mu=j_\mu^{(1)}+\xi[A_\mu+2\,\partial_\mu\Im s],
\eeq
with $j_\mu^{(1)}$ given in eq.~(\ref{WZ28}).
We also have that 
\beq
\label{FI17}
\Im f_{\widehat X} = \Im f_{X^{(1)}}
- {2\over3}\xi\,\partial^\mu [A_\mu + 2\,\partial_\mu\Im s],
\eeq
and then the equation
\beq
\label{FI18}
\partial^\mu \widehat j_\mu=-{3\over 2}\Im f_{\widehat X}
\eeq
is just the equation for the (non-)conservation of $j_\mu^{(1)}$.  
Similarly, the energy-momentum tensor in supercurrent superfield 
(\ref{FI15}) is
\beq
\label{FI19}
\widehat T_{\mu\nu}=T_{\mu\nu}^{(1)}-{\xi\over 3}(\partial_\mu\partial_\nu-\eta_{\mu\nu}\Box)(C+2\Re s).
\eeq
Since
\beq
\label{FI20}
\Re f_{\widehat X}=\Re f_{X^{(1)}}-{2\over 3}\xi\, \Bigl[d-\Box(C+2\Re s) \Bigr],
\eeq
the equation
\beq
\label{FI21}
\widehat T^\mu{}_\mu={3\over 2}\Re f_{\widehat X}
\eeq
is equivalent to
\beq
\label{FI22}
T^{(1)\mu}{}_\mu={3\over 2}\Re f_{X^{(1)}}-\xi d
\eeq 
and simply determines the trace of $T_{\mu\nu}^{(1)}$ as in eq.~(\ref{WZ27b}), but in the 
presence of the Fayet-Iliopoulos term. All physically significant
equations are unchanged by the introduction of $S$, but all quantities are now formally 
gauge-invariant.

Hence, we have needed $4_B+4_F$ new degrees of freedom to obtain the
gauge-invariant Ferrara-Zumino structure (\ref{FI15}). 
But the introduction of $S$ is 
purely formal, as is the problem of the gauge variation of the superfield
FI term. It does not 
alter the dynamics of the theory nor does it change its symmetry properties. 
The new fields do not propagate.\footnote{Using a non-dynamical superfield to restore
a superspace local symmetry has been introduced in section 6 of ref.~\cite{HKLR}, to
gauge symmetries leaving the K\"ahler potential invariant up to a K\"ahler transformation.}

From the point of view of global supersymmetry, this is a fully satisfactory
formulation of the supercurrent structure with a FI term. 
The introduction of the $S$ field, which as earlier observed can also be viewed 
as a particular K\"ahler transformation, is independent from the form of the superpotential: 
K\"ahler invariance of the globally supersymmetric theory 
does not involve the superpotential. Supergravity is different, it can be  
expressed in terms of the function ${\cal G} = K + \ln (W\ov W)$, and a
K\"ahler transformation must be compensated by a superpotential 
transformation.

\subsection{On the Fayet-Iliopoulos term in supergravity}

We now wish to uplift the construction of the previous section to supergravity. Our goal
is to show how does the introduction of the chiral superfield $S$ survive and is actually naturally 
included in the supergravity coupling. We use the superconformal formulation of Poincar\'e
supergravity, with a compensating supermultiplet to gauge fix superconformal symmetries 
absent in the super-Poincar\'e algebra.
In the minimal sets of supergravity auxiliary fields, the compensating multiplet is either chiral 
({\it old minimal}) or real linear ({\it new minimal}). We begin with the new minimal construction.

In the superconformal formulation, new minimal supergravity has Lagrangian
\beq
\label{sg1}
{\cal L}_{n.m.} = {3\over2}\left[ L \ln \left({L\over S_0\ov S_0}\right) - L  \right]_D , 
\eeq
where $L$ is the compensating real linear multiplet (Weyl weight $w=2$), 
$S_0$ is chiral with Weyl and chiral weights $w=n=1$ and $[\ldots]_D$ denotes the invariant
$D$--density formula\footnote{We refer to Kugo and Uehara \cite{KU} for the 
superconformal calculus. We (almost) use their conventions and notations.}.
Since $[ L ( \Lambda+\ov\Lambda)]_D$ is a derivative for any chiral $\Lambda$, 
$S_0$ contributes to the Lagrangian by a derivative only: it does not generate any field equation 
and its role is only to give 
the correct dimension to the argument of the logarithm. The theory has the obvious gauge invariance 
\beq
\label{sg2}
S_0 \quad\longrightarrow\quad e^{-\Lambda'} S_0\,,
\eeq
which allows to choose a gauge where $S_0$ is a nonzero dimensionful constant.
Tensor calculus then shows that\footnote{$C_L$ denotes the lowest component of the linear multiplet $L$.}
\beq
\label{sg3}
e^{-1} {\cal L}_{n.m.} = - {1\over2} C_L\, R - {3\over2} \ln \left({C_L\over z_0\ov z_0}
\right) \Box C_L + \ldots
\eeq
and the gauge-fixing of dilatation symmetry corresponds to
\beq
\label{sg4}
C_L = \kappa^{-2}\equiv \frac{M_{P}^2}{8\pi}
\eeq
in the Einstein frame.

We can then couple new minimal supergravity to gauge and matter supermultiplets. Assuming vanishing Weyl weights for the latter, the compensating 
multiplet $L$ is used to obtain the required Weyl weight $w=2$ in $D$--densities while $S_0$ can be 
used in the superpotential $F$--density provided the gauge invariance (\ref{sg2}) is preserved. 
Consider then the theory 
\beq
\label{sg5}
\begin{array}{rcl}
{\cal L}_{n.m.} &=& \displaystyle 
{3\over2}\left[  L \ln \left({L\over S_0\ov S_0}\right) - L  
+{1\over3}L K(\ov\phi_i e^{q_i A} \phi_i ) + {2\over3}\hat\xi L(A+S+\ov S)\right]_D
\crbig
&&\displaystyle
+ \Bigl[ S_0^3 \, W(\phi_i)  + {1\over4} g(\phi_i){\cal W}{\cal W} \Bigr]_F.
\end{array}
\eeq
There are two distinct cases. Firstly, if the superpotential does not vanish, 
gauge transformation (\ref{sg2}) must then
be compensated by a transformation of matter superfields
\beq
\label{sg2b}
\phi_i  \quad\longrightarrow\quad e^{Q_i\Lambda'}\phi_i 
\eeq 
with invariance conditions $W(e^{Q_i\Lambda'}\phi_i) = e^{3\Lambda'}W(\phi_i)$ 
and $g(e^{Q_i\Lambda'}\phi_i)=g(\phi_i)$. It is an $R$--symmetry acting on the $\phi_i$ and 
invariance of the K\"ahler potential $K$ implies then that $A$ gauges this $R$--symmetry, up to, 
maybe, a non--$R$ global symmetry.
Notice that this condition is not related to the FI term: new minimal supergravity only admits 
$R$--symmetric superpotentials. Recall also that in the $R$--symmetric case, the natural
supercurrent structure (\ref{FI8}) of the globally supersymmetric theory has
$$
X = 4W = \ov{DD} \, {1\over3}\sum_i Q_i \phi_i K_{\phi_i}
$$
using the first field equation (\ref{FI6}). Hence, an improvement transformation (\ref{impr2}) can cancel $X$
and we expect that the global theory admits a coupling to new minimal supergravity.

If the superpotential vanishes, gauge transformations of $S_0$ and $\phi_i$ remain decoupled.
Since  theory (\ref{sg5}) only depends on $S_0 \exp(-{2\over3}\hat \xi S)$ in a derivative term, 
it is clear that $S$ can either be produced from $S_0$ by a gauge transformation  (\ref{sg2}) or
absorbed into $S_0$ by the inverse transformation. In any case, gauge invariances of $A$ and $S_0$
are preserved and independent. 
In the gauge $S=0$ however, gauge variations of $S_0$ and $A$ are 
identified, the quantity $\ov S_0 e^{-{2\over3}\hat \xi A}S_0$ is invariant:
the FI coefficient is then related to the charge of $S_0$ under the 
transformation (\ref{sg2}) gauged by $A$. 

The old minimal formulation of theory (\ref{sg5}) is obtained as follows. We first replace 
the linear $L$ by an unconstrained real vector superfield $U$ with Weyl weight $w=2$:
\beq
\label{sg6}
\begin{array}{rcl}
{\cal L}_{n.m.} &=& \displaystyle 
{3\over2}\left[  U \ln \left({U\over S_0\ov S_0}\right) - U 
+{1\over3}U K(\ov\phi_i e^{q_i A} \phi_i ) + {2\over3}\hat\xi U(A+S+\ov S)\right]_D
\crbig
&&\displaystyle
+ \Bigl[ S_0^3 \, W(\phi_i)  + {1\over4} g(\phi_i){\cal W}{\cal W} \Bigr]_F .
\end{array}
\eeq
Now $S$ induces an algebraic field equation, with solution $U=L$. 
But we instead eliminate $U$ by solving its field equation
\beq
\label{sg7}
U = S_0\ov S_0 \exp\left( -{1\over3}K - {2\over3} \hat\xi(A+S+\ov S) \right).
\eeq
The resulting theory is
\beq
\label{sg8}
{\cal L}_{o.m.} = - {3\over2} \left[ S_0\ov S_0 \exp\left( -{1\over3}K 
- {2\over3} \hat\xi(A+S+\ov S) \right) \right]_D
+ \Bigl[ S_0^3 \, W + {1\over4} g \,{\cal W}{\cal W} \Bigr]_F 
\eeq
and $S_0$ is now the chiral compensating multiplet of the old minimal formalism.
Notice that when the linear $L$ is replaced by the real $U$, gauge invariance is obtained 
by assigning to $S$ a gauge variation such that $U$ is invariant. 
A combined gauge transformation of $S_0$, $S$ and (if $W$ is not zero and $R$--symmetric)
$\phi_i$ allows then to eliminate $S$ which does not play any dynamical role, in complete 
correspondence with the globally supersymmetric case.

In old minimal theory (\ref{sg8}), tensor calculus indicates that
\beq
\label{sg9}
e^{-1}{\cal L}_{o.m.} = - {1\over2} z_0\ov z_0 \exp\left( -{1\over3}K \right) \, R 
+{1\over 2} \,\hat\xi\, z_0\ov z_0 e^{-K/3} \, d
+ \ldots
\eeq
The dilatation fixing condition is now
\beq
\label{sg10}
z_0\ov z_0 \exp\left( -{1\over3}K \right) = {1\over\kappa^2},
\eeq
in the $S=0$ gauge, 
which, in the Wess-Zumino gauge, is also the lowest component of eq.~(\ref{sg7}), 
with $U_{\theta=0}=C = \kappa^{-2}$. In both 
theories (\ref{sg5}) and (\ref{sg8}) the FI coefficient is
\beq
\label{sg11}
\xi = {\hat\xi\over\kappa^2} . 
\eeq

With a superpotential breaking any $R$--symmetry, the natural supercurrent structure (\ref{FI8})
always has a nonzero anomaly superfield $X$. The theory cannot be coupled to new minimal
supergravity without breaking gauge invariance (\ref{sg2}) which allows to gauge 
$S_0$ away. 
In the dual, old minimal version (\ref{sg8}), one can for instance eliminate the $S$ field from the
$D$--density using a K\"ahler transformation, or a holomorphic field redefinition of $S_0$.
The result would be
\beq
\label{sg12}
{\cal L}_{o.m.} = - {3\over2} \left[ S_0\ov S_0 \exp\left( -{1\over3}K 
- {2\over3} \hat\xi A \right) \right]_D
+ \Bigl[ S_0^3 \, e^{2\hat\xi S} W + {1\over4} g \,{\cal W}{\cal W} \Bigr]_F .
\eeq
The field equation for $S$ is now a constraint imposing $W=0$. Hence, further terms in $S$ are needed
to make the field equations consistent. One should then consider a system with a supplementary dynamical
chiral superfield $S$, {\it i.e.} with a non-minimal supergravity field content with $16_B+16_F$ off-shell fields.
As observed by Komargodski and Seiberg \cite{KS}, this is the natural supergravity formulation 
for the supercurrent structure with $X\ne0\ne\chi_\alpha$. This formulation of supergravity is 
usually named $16+16$ supergravity \cite{1616, Lang, Siegel, ADO}.

One easily checks that in the global supersymmetry limit, the constraint induced by $S$ in 
the supergravity theory (\ref{sg12}) vanishes as $\kappa^2\rightarrow 0$. Hence the natural supercurrent structure (\ref{FI8}) is valid for an arbitrary superpotential. 

It is clear that in the old or new minimal superconformal setups, one scale only is 
generated by the gauge-fixing of dilatations. Hence the resulting FI term is bound to be 
proportional to $\kappa^{-2}$ and the
rigid limit $\kappa^{-2}\rightarrow\infty$ must be taken by requiring $\hat\xi\kappa^{-2}$ finite.
In the $16+16$ supergravity setup however, with a linear $L$ and a chiral $S_0$, 
one can easily generate two independent scales at the price of $4_B+4_F$ 
supplementary dynamical fields. For instance, 
$$
-{3\over2}\Bigl[ S_0\ov S_0 e^{-K/3} \Bigr]_D
$$
includes an Einstein term while
$$
\xi[ LA ]_D
$$
includes a ``Fayet-Iliopoulos field" but no Einstein term. We can then associate the FI coefficient $\xi$ to
the background value $\langle C_L\rangle$ while $S_0$ is gauge-fixed as in eq.~(\ref{sg10}) and generates 
then the Planck scale. Assuming $\langle C_L\rangle \ll \kappa^{-2}$ leads to an effective 
Fayet-Iliopoulos term with scale independent from the Planck scale in the global supersymmetry limit.

\section{Conclusions}
\setcounter{equation}{0}

We have reviewed the supercurrent structure of $N=1$ supersymmetric gauge theories in terms of the supercurrent and anomaly superfields. We have shown how superfield identities together with the field equations lead to a ``natural'' supercurrent structure for an arbitrary gauge-invariant Wess-Zumino model.  We have pointed out that the supercurrent superfield is written in terms of off-shell superfields of the theory, but the supercurrent equation only holds on-shell. Then, the interpretation of the components of the supercurrent superfield as Noether currents requires the field equations. This is explicitly visible for the auxiliary field contribution to the energy-momentum tensor.

The natural supercurrent structure turns out to contain the Belinfante improved canonical (Noether) energy-momentum tensor and the Noether current for $U(1)_{\widetilde R}$ transformations, which leave chiral superfields inert. We have discussed two transformations of the natural supercurrent structure, which induce improvements of the supercurrent and energy-momentum tensor. The first one leads to a Ferrara-Zumino structure, which, for a generic theory, neither contains the new improved energy-momentum tensor of CCJ nor the Noether current of $U(1)_R$ transformations under which chiral superfields have a particular $R$--charge. The supercurrent structure containing these currents is obtained by a different transformation which in general retains the $16_B+16_F$ structure. Both transformations coincide for scale-invariant K\"ahler potentials.

Our unambiguous procedure to find supercurrent structures also applies to a supersymmetric gauge theory with a Fayet-Iliopoulos term, providing us with a natural gauge-invariant $16_B+16_F$ supercurrent structure. We have presented a possibility to transform the latter into a gauge-invariant Ferrara-Zumino structure. This relies on the introduction of a non-dynamical chiral superfield $S$, which does not change the field equations and the symmetries of the theory. It is just a purely formal device to obtain gauge-invariant Lagrangian superfields. We have then coupled the theory to supergravity and have pointed out that $S$ is naturally present due to a gauge transformation of the chiral multiplet $S_0$ in the superconformal formulation of new minimal supergravity. After performing the duality transformation to the old minimal formulation, we have found that the equations of motion for $S$ impose a vanishing superpotential. These complications vanish in the $\kappa^2\rightarrow 0$ limit to global supersymmetry, where $S$ can be always non-dynamical. For $R$--invariant superpotentials, they can be avoided by gauging the $R$--symmetry, which allows to gauge away $S$ and its equations of motion. For a generic superpotential this is not the case anymore and one is forced to introduce kinetic terms for $S$. This then turns $S$ from a mere formal tool to $4_B+4_F$ additional propagating fields, in agreement with the results of ref.~\cite{KS}.

\section*{Acknowledgements}

We wish to thank Nicola Ambrosetti and Matthias Blau for useful 
conversations and contributions. J.-P. D. thanks the Simons Center 
for Geometry and Physics in Stony Brook for hospitality during part of this work. 
This research has been supported by the Swiss National Science Foundation. 

\renewcommand{\thesection}{\Alph{section}}
\setcounter{section}{0}
\setcounter{equation}{0}
\renewcommand{\theequation}{\thesection.\arabic{equation}}
\section{Supersymmetric improvement transformation}

We take the following expansion of the supercurrent superfield $J_\mu$:
\beq
\begin{array}{rcl}
J_\mu (x,\theta,\ov\theta) &=& {8\over 3}j_\mu(x) 
+ \theta (S_\mu + 2\sqrt2\, \sigma_\mu\ov\psi_X )
+ \ov\theta (\ov S_\mu - 2\sqrt2\, \ov\sigma_\mu\psi_X )
\crbig
&& - 2i \,\theta\theta\,\partial_\mu\ov x + 2i \,\ov{\theta\theta}\, \partial_\mu x
\crbig
&& \displaystyle
+ \theta\sigma^\nu\ov\theta\left[ 8\,T_{\mu\nu} - 4\,\eta_{\mu\nu}\Re f_X
- {1\over 2}\epsilon_{\mu\nu\rho\sigma}\left({8\over 3}\, \partial^\rho j^\sigma - F^{\rho\sigma} \right)\right]
\crbig
&& \displaystyle - \frac{i}{2} \theta\theta\ov\theta ( \partial_\nu S_\mu \sigma^\nu 
+ 2 \sqrt2\, \ov\sigma_\mu \sigma^\nu \partial_\nu\ov\psi_X)
\crbig
&& \displaystyle + \frac{i}{2} \ov{\theta\theta}\theta ( \sigma^\nu \partial_\nu \ov S_\mu
+ 2\sqrt2\, \sigma_\mu \ov\sigma^\nu \partial_\nu\psi_X)
\crbig
&& \displaystyle
- \frac{2}{3}\theta\theta\ov{\theta\theta}\, \Bigl(  2 \, \partial_\mu \partial^\nu j_\nu
- \Box j_\mu \Bigr).
\end{array}
\eeq
The components of the superfields $X$ and $\chi_\alpha$ read
\beq
\begin{array}{rcl}
 X &=& x + \sqrt2\,\theta\psi_X - \theta\theta\, f_X-i\theta\sigma^\mu\bar\theta\partial_\mu x-\frac{i}{\sqrt{2}}\theta\theta\bar\theta\bar\sigma^\mu\partial_\mu\psi_X-\frac{1}{4}\theta\theta\ov{\theta\theta}\Box x ,
 \crbig
 \chi_\alpha &=& -i \lambda_\alpha + \theta_\alpha \, D + \frac{i}{2}(\theta\sigma^\mu\ov\sigma^\nu)_\alpha F_{\mu\nu} - \theta\sigma^\mu\bar\theta\partial_\mu\lambda_\alpha-\theta\theta(\sigma^\mu\partial_\mu\ov\lambda)_\alpha\nonumber
 \crbig
 &&- \frac{1}{2}\theta\theta(\sigma^\mu\bar\theta)_\alpha(\partial_\nu F^\nu{}_\mu-i\partial_\mu D)+\frac{i}{4}\theta\theta\ov{\theta\theta}\Box\lambda_\alpha,
\end{array}
\eeq
with $F_{\mu\nu}=-F_{\nu\mu}$ and $\partial_{[\mu}F_{\nu\rho]}=0$.
If the real superfield $\cal G$ of the transformation (\ref{impr2}) has the expansion
\beq
\begin{array}{rcl}
 \cal G &=& C_g+i\theta\chi_g-i\bar\theta\bar\chi_g+\theta\sigma^\mu\bar\theta v_{g\mu}+\frac{i}{2}\theta\theta(M_g+iN_g)-\frac{i}{2}\ov{\theta\theta}(M_g-iN_g)
\crbig
&& + i\theta\theta\bar\theta(\bar\lambda_g+\frac{i}{2}\partial_\mu\chi_g\sigma^\mu)-i\ov{\theta\theta}\theta(\lambda_g-\frac{i}{2}\sigma^\mu\partial_\mu\bar\chi_g)+\frac{1}{2}\theta\theta\ov{\theta\theta}(D_g-\frac{1}{2}\Box C_g),
\end{array}
\eeq
then the components of the transformed superfields $\widetilde J_\mu$, $\widetilde X$ and $\widetilde\chi_\alpha$ read
\beq
\begin{array}{rcl}
\widetilde j_\mu &=& j_\mu-3v_{g\mu} ,
\crbig
\widetilde S_\mu &=& S_\mu+8\sigma_{[\mu}\bar\sigma_{\nu]}\partial^\nu\chi_g,
\crbig
\widetilde\psi_X &=& \psi_X+2\sqrt{2}i\lambda_g+2\sqrt{2}\sigma^\mu\partial_\mu\bar\chi_g,
\crbig
\widetilde x &=& x+2i(M_g-iN_g),
\crbig
\widetilde T_{\mu\nu} &=& T_{\mu\nu}+(\partial_\mu\partial_\nu-\eta_{\mu\nu}\Box) C_g,
\crbig
\widetilde f_X &=& f_X+2D_g-2\Box C_g+2i\partial_\mu v^\mu_g,
\crbig
\widetilde F_{\mu\nu} &=& F_{\mu\nu}-24\partial_{[\mu}v_{g\nu]},
\crbig
\widetilde\lambda &=& \lambda-12\lambda_g,
\crbig
\widetilde D &=& D-12D_g.
\end{array}
\eeq

\end{document}